\documentclass[prl,a4paper,nofootinbib,
twocolumn,
]{revtex4-1}
\usepackage[T1]{fontenc}
\usepackage{graphicx} 
\usepackage{bbold}
\usepackage{mathtools}
\usepackage{amsmath}
\usepackage{amsfonts}
\usepackage{amssymb}
\usepackage{slashed} 
\usepackage{color}
\usepackage[table]{xcolor}
\usepackage{hyperref}
\usepackage{bm}% bold math
\usepackage{mathrsfs}
\def\eq#1{{Eq.~(\ref{#1})}}

\def\Tr{\mbox{Tr}\,}

\def\etal{{\it et al.}}

\colorlet{grayline}{gray!70}
\definecolor{mf}{rgb}{204,0,0}
\definecolor{blueline}{rgb}{0,0.27,0.55}
\definecolor{DarkGray}{gray}{0.4}
\definecolor{Gray}{gray}{0.6}
\definecolor{oucrimsonred}{rgb}{0.6, 0.0, 0.0}
\definecolor{persianblue}{rgb}{0.11, 0.22, 0.73}
\definecolor{forestgreen}{rgb}{0.13,0.35,0.13}
 \hypersetup{colorlinks, citecolor=forestgreen, linkcolor=forestgreen, urlcolor=forestgreen}
\def\hhref#1{\href{http://arxiv.org/abs/#1}{#1}} % in bibliography
\newcommand{\be}{\begin{equation}}
\newcommand{\ee}{\end{equation}}
\newcommand{\bea}{\begin{eqnarray}}
\newcommand{\eea}{\end{eqnarray}}
\newcommand{\nn}{\nonumber}

\newcommand*\xbar[1]{%
  \hbox{\;%
    \vbox{%
      \hrule height 0.5pt % The actual bar
      \kern0.5ex%         % Distance between bar and symbol
      \hbox{%
        \kern-0.25em%      % Shortening on the left side
        \ensuremath{#1}%
        \kern-0.07em%      % Shortening on the right side
      }%
    }%
  }%
} 
\newcommand{\com}[1]{}
\newcommand{\gsim}{\lower.7ex\hbox{$\;\stackrel{\textstyle>}{\sim}\;$}}
\newcommand{\lsim}{\lower.7ex\hbox{$\;\stackrel{\textstyle<}{\sim}\;$}} 

\newcommand{\bc}{\begin{center}}
\newcommand{\ec}{\end{center}}

\newcommand{\K}{K^{*}(892)^0}

\font\beeg=cmr17 scaled 1800
\newbox\ibox
\def\versal#1{\setbox\ibox=\hbox{{\beeg #1}~}%
	    \noindent\global\hangindent=\wd\ibox\global\hangafter-2%
	    \sc\smash{\llap {\lower 14pt \box\ibox}}}
\begin{document}
\title[]{ \Large \color{oucrimsonred} \textbf{ 
 Bell inequality is violated
  in  $B^0\to J/\psi \, \K$ decays
}}
\author{\bf M. Fabbrichesi$^{a}$}
\author{\bf   R. Floreanini$^{a}$}
\author{\bf E. Gabrielli$^{{b,a,c,d}}$ and} 
\author{\bf L. Marzola$^{{d}}$}

\affiliation{$^{a}$ INFN, Sezione di Trieste, Via Valerio 2, I-34127 Trieste, Italy}
\affiliation{$^{b}$ Physics Department, University of Trieste, Strada Costiera 11, I-34151 Trieste, Italy}
 \affiliation{$^{c}$ CERN, Theoretical Physics Department, Geneva, Switzerland}
\affiliation{$^{d}$ Laboratory of High-Energy and Computational Physics, NICPB, R\"avala 10,  10143 Tallinn, Estonia}

 \begin{abstract}
   \noindent The violation of  the Bell inequality is one of the  hallmarks of quantum mechanics and can be used to rule out  local deterministic alternative descriptions. We utilize  the  data analysis published by the LHCb collaboration  on the helicity amplitudes for  the decay $B^0\to J/\psi \,\K$    to compute the entanglement among the polarizations of the final vector mesons and  the violation of  the Bell inequality that it entails. We find that quantum entanglement can be detected with a significance well above  5$\sigma$ (nominally  84$\sigma$) and Bell inequality is violated with a significance well above  5$\sigma$
(nominally 36$\sigma$)---thereby firmly establishing these distinguishing feature of quantum mechanics at   high energies in a collider setting and in the presence of strong and weak interactions. Entanglement is also present and the Bell inequality  is violated    in other decays of the $B$ mesons into vector mesons, but with lesser significance.
 \end{abstract}

\maketitle

%%%%%%%%%%%%%%%%%%%%%%%%%%%%%%%%%%%%%%%%%%%%%%%%%%%%%%%%%%%%%%%%%%%
%\newpage
%\pagestyle{plain}
%\tableofcontents

{\versal  Introduction.---} The violation of the Bell inequality~\cite{Bell} is a phenomenon that shows that quantum mechanics cannot be explained by any local hidden variable theory, which assumes that physical systems have definite properties independent of measurement.
It has been verified experimentally with the polarizations of low-energy (that is, few eV) photons in~\cite{Aspect:1982fx,Weihs:1998gy}:
two photons are prepared into a singlet state  and their polarizations  measured along different directions to verify their entanglement~\cite{Horodecki:2009zz} and the  violation of  Bell inequality.

Verifying quantum entanglement and the violation of the Bell inequality in the presence of  strong and weak interactions would tell us whether these fundamental forces of nature  exhibit quantum correlations and non-locality, which would have profound implications for our understanding of reality. 
In order to  test the inequality at higher energies, we need a sufficiently heavy scalar (or pseudo-scalar) particle decaying  into two spin-1/2 or spin-1 states.  While we  do not know of  any data in the case of fermions, these are available  for  the  two final states being  massive  vector-like particles. The set-up in the latter case closely resembles that in which the polarizations of two photons prepared into a singlet state are measured---except that the photon polarizations are described by a two-value quantum state, or qubit, while  those of the  massive spin-1 state have three values and are described by  \textit{qutrits}. 

The most promising examples can be found among $B$ mesons decaying into comparatively heavy final states with  approximately equal shares of longitudinal and transverse polarizations. In addition,  larger branching fractions make for better statistics.
These requirements single out the decay   $B^0\to J/\psi \K$ as the best candidate. 
The data analysis of the LHCb collaboration for this decay~\cite{LHCb:2013vga} provides the helicity amplitudes necessary for the test. They make it possible to extend the testing of the  violation of Bell identities to energies of the order of 5 GeV---which are a billion times larger than those  utilized in~\cite{Aspect:1982fx,Weihs:1998gy,Hensen:2015ccp}. 
The same decay has previously been studied by the experiments CLEO~\cite{CLEO:1997ilq}, CDF~\cite{CDF:2000edf}, Belle~\cite{Belle:2005qtf}, BaBar~\cite{BaBar:2007rbr}, and D0~\cite{D0:2008nly}. We only use the most recent analysis  because it is the most precise.

In this work, we  explain why the $B^0\to J/\psi\, \K$  decay provides a  most favorable setting, introduce two operators to quantify entanglement and violation of the Bell inequality for a two-qutrit system, compute the expectation values of these two operators   using the polarization amplitudes provided in \cite{LHCb:2013vga}  and show that quantum entanglement is present with a significance
well above 5$\sigma$ (nominally  84$\sigma$) and the Bell inequality is violated with a significance well above 5$\sigma$ (nominally  36$\sigma$). This result  firmly establishes  this quantum mechanical hallmark for a  system of two qutrits, and it does it at high energies and in the presence of strong and weak interactions---thereby extending what is known to be true  for qubits, at low energies and for  electromagnetic interactions.

We also analyze the decays   $B^0\to \phi \,\K$, $B^0\to \rho \,\K$,  $B_s\to J/\psi\, \,\phi$ and $B_s\to \phi \,\phi$,  which have sizable transverse polarizations, and find that quantum entanglement is present with a significance of
5.35$\sigma$, 5.84$\sigma$,  22.8$\sigma$, and 19.8$\sigma$, respectively,  and they  violate   the Bell inequality with a significance of 1.1$\sigma$, 1.4$\sigma$,  5.8$\sigma$ and 8.2$\sigma$, respectively.
 
Previous inquiries about Bell inequality violations with data from heavy-particle physics have been presented for kaons~\cite{Benatti2} and for the $B^0$-$\bar B^{\, 0}$ system~\cite{Go:2003tx}. Both of these examples, though providing  important clues, are indirect  tests:  the first relies on the measure of the  CP violating parameter $\varepsilon^{\prime}/\varepsilon$, the second on  oscillations in flavor space.

\vskip0.5cm
{\versal Materials.---} The  analysis of the decay $B^0\to J/\psi \, \K$ in \cite{LHCb:2013vga} is based on the   data sample collected in $p p$ collisions at 7 TeV (part of run 1 of the LHC) with the LHCb detector and corresponds to an integrated luminosity of 1 fb$^{-1}$. The branching fraction for this decay is $(1.27\pm0.05)\times 10^{-3}$~\cite{ParticleDataGroup:2022pth}.

 The selection of $B^{0}\to J/\psi \,\K$ events, as explained in \cite{LHCb:2013vga}, is based upon the combined decays of the $J/\psi \to \mu^{+}\mu^{-}$ and the $\K  \to K^{+}\pi^{-}$ final states. The muons, as they  leave two oppositely-charged tracks originating from a common vertex, are selected by taking their  transverse momentum  $p_{T} > 500$ MeV/c.   The invariant mass of this pair of muons is required to  be in the  range between 3030 and 3150 MeV$/c^{2}$. 
The kaon and the pion leave two oppositely-charged tracks that originate from the same vertex. It is required that the $\K$  has transverse momentum $p_{T} > 2$ GeV/c and invariant mass in the  range 826-966 MeV/c$^{2}$. The $B^{0}$  are reconstructed from the $J/\psi$ and $\K$ candidates, with the invariant mass of the $\mu^{+}\mu^{-}$ pair constrained to the $J/\psi$ mass. The resulting $B^{0}$ candidates are required to have an invariant mass of the system $J/\psi K^{+}\pi^{-}$  in the range 5150-5400 MeV/c$^{2}$.

The polarizations of the spin-1 massive particles  $J/\psi$ and  $\K$ can be reconstructed  using the momenta of the final charged mesons and  leptons in which they decay. 
The differential decay rate is described in terms of three angles: two  angles are defined by the direction of the $\mu^{+}$ momentum with respect to  the $z$ and $x$ axes in the $J/\psi$ rest frame, and one by the direction of the momentum  of the $K^{+}$ with respect to  the  opposite direction of the momentum of the $J/\psi$  in the $\K \to K^{+}\pi^{-}$ rest frame, as shown in  Fig. (2) of  \cite{LHCb:2013vga}.
The  longitudinal polarization amplitudes $A_{0}$ and the two transverse amplitudes $A_{\perp}$ and $A_{\parallel}$ are found  as coefficients of combinations of trigonometric functions of these three angles~\cite{Dighe:1995pd}.

The analysis in~\cite{LHCb:2013vga} gives the two complex polarization amplitudes $A_{\parallel}$ and $A_{\perp}$ as well as the non-resonant  amplitude $A_s$.
We need only the former two and take the following values for the squared moduli and phases of these polarization amplitudes:
\begin{align}
|A_{\parallel}|^{2} &=  0.227 \pm 0.004\; \text{(stat)} \pm 0.011\; \text{(sys)} \nn \\
|A_{\perp}|^{2} &=  0.201 \pm 0.004\; \text{(stat)}  \pm 0.008\; \text{(sys)} \nn \\
\delta_{\parallel} \; [\text{rad}] &=  -2.94 \pm 0.02\; \text{(stat)} \pm 0.03\; \text{(sys)} \nn \\
\delta_{\perp}  \;  [\text{rad}] &= 2.94 \pm 0.02\; \text{(stat)}  \pm 0.02\; \text{(sys)}  \, ,\label{data0}
\end{align}
with $|A_{0}|^{2}+|A_{\perp}|^{2}+|A_{\parallel}|^{2}=1$, and we can take $\delta_0=0$ because there are only two physical phases.
The correlations among the amplitude and phase uncertainties are also provided in~\cite{LHCb:2013vga}. The polarization amplitudes are complex mostly because of the final-state interactions (see, for instance, \cite{Beneke:1999br}). The values in \eq{data0} have errors that are 2 or 3 times smaller than those of the previous analyses~\cite{CLEO:1997ilq,CDF:2000edf,Belle:2005qtf,BaBar:2007rbr,D0:2008nly}.

The decays of the $J/\psi$ and $K^*$ take place well outside of the range of the strong interactions ongoing at the time of their production (which is due to gluons exchange and is about $3 \times 10^{-5}$ fm~\cite{Yamamoto:2008ze}) as well as of the final-state interactions. The distance between the two mesons, at the time they both have decayed, can be estimated to be $d \simeq 1.1 \times 10^3 $ fm. This distance must be compared with the typical range of the virtual meson exchange, that is at most equal to $\lambda_{\pi} =$ 1.5 fm. We thus obtain that $d/\lambda_{\pi} \simeq 750$, indicating the impossibility of any strong interaction exchange between the two decaying particles.
About the same distance is found for the decay into $J/\psi \phi$, while values of $d$ between 100 and 10 are found for the other decays in Table~\ref{tab:others}, namely $\phi \phi$, $\phi K^*$, and, with the least separation, $\rho K^*$.

\vskip0.5cm
{\versal Methods.---} There are three helicity amplitudes for the decay of a scalar, or pseudo-scalar,  into two massive spin-1 particles:
\be
h_{\lambda} = \langle V_{1}(\lambda) V_{2}(-\lambda)| {\cal H} |B\rangle\quad \text{with} \quad \lambda=(+,\, 0,\,-)\, ,
\ee
and ${\cal H}$ is the interaction Hamiltonian  giving rise to the decay.
For the spin quantization axis ($\hat z$) we use the direction of the momenta of the decay products in the $B^0$ rest frame.
Helicities are here defined with respect to the $\hat z$ direction  in the rest frame of one of the two  spin-1 particles and $(+,\, 0,\,-)$ is a shorthand  for $ (+1,\, 0,\,-1)$.

The polarizations in the decay  are described by a quantum state that is pure for any values of the helicity amplitudes~\cite{Fabbrichesi:2023cev,Fabbrichesi:2023jep}. This state can be written as
\bea
|\Psi \rangle &=& \frac{1}{\sqrt{|H|^2}} \Big[  h_+\, |V_{1}(+)  V_{2}(-)\rangle \nn \\
& +&  h_0 \, |V_{1}(0)  V_{2}(0)\rangle+  h_-\, |V_{1}(-)  V_{2}(+)\rangle \Big] \, ,\label{pure}
\eea
with
\be
|H|^2= |h_0|^2 + |h_+|^2 + |h_-|^2 \, .
\ee
The relative weight of the transverse components $ |V_{1}(+)  V_{2}(-)\rangle $ and $ |V_{1}(-)  V_{2}(+)\rangle $ with respect to the longitudinal one $| V_{1}(0)  V_{2}(0 )\rangle$ is controlled  by the conservation of angular momentum. In general, only the helicity is conserved and the state in \eq{pure} 
belongs to the $J_z=0$ component of the $S=0,1$ or $2$ states.

The polarization density matrix $\rho = |\Psi \rangle \langle \Psi |$ can be written in terms of the helicity amplitudes as
\be
\small
\rho=  \frac{1}{|H|^2}\, \begin{pmatrix} 
  0 & 0 & 0 & 0 & 0 & 0 & 0 & 0 & 0  \\
  0 & 0 & 0 & 0 & 0 & 0 & 0 & 0 & 0  \\
  0 & 0 &  h_+ h_+^* & 0 &  h_+  h_0^*& 0 &  h_+ h_-^*& 0 & 0  \\
  0 & 0 & 0 & 0 & 0 & 0 & 0 & 0 & 0  \\
  0 & 0 & h_0 h_+^* & 0 & h_0  h_0^*  & 0 & h_0 h_-^*& 0 & 0  \\
  0 & 0 & 0 & 0 & 0 & 0 & 0 & 0 & 0  \\
  0 & 0 &  h_- h _+^*& 0 &  h_- h_0^*& 0 &   h_-h_-^*& 0 & 0  \\
  0 & 0 & 0 & 0 & 0 & 0 & 0 & 0 & 0  \\
  0 & 0 & 0 & 0 & 0 & 0 & 0 & 0 & 0  \\
\end{pmatrix} \, ,
\label{rhoBVV}
\ee
on the basis given by the tensor product of the polarizations $(+,\, 0,\, -)$ of the produced spin-1 particles.
%, defined as the eigenstates of the  spin operator in the direction $\hat z$ with eigenvalues $\{1,0,-1\}$.

The helicity amplitudes are mapped into the polarization amplitudes used in \eq{data0} by the correspondence
\be
\frac{h_{0}}{|H|}= A_{0} \,, \quad \frac{h_{+}}{|H|}  = \frac{ A_{\parallel}+A_{\perp}}{\sqrt{2}}\, , \quad
\frac{h_{-}}{|H|}  = \frac{ A_{\parallel} -A_{\perp}}{\sqrt{2}} \, .
\ee

Having written the  density matrix,
we can study the entanglement among the polarizations of the two massive vector particles by means of a simple observable. For a bipartite pure state, like the one in \eq{rhoBVV}, the von Neumann entropy~\cite{Horodecki:2009zz}
\be
\mathscr{E} = - \Tr[\rho_{S_{A}} \ln \rho_{S_{A}}] =  - \Tr[\rho_{S_{B}} \ln \rho_{S_{B}}] \, , \label{E}
\ee
quantifies entanglement; in \eq{E}, $\rho_{S_{A}}$ and $\rho_{S_{B}}$ are the reduced density matrices for the two sub-systems $S_{A}$ and $S_{B}$, which are the two spin-1 mesons in the decay under consideration.
The von Neumann entropy of a two-qutrit system satisfies  $0\leq \mathscr{E} \leq \ln 3$. The first equality is true if and only if the bipartite state is separable, the second if the bipartite state is maximally entangled.

The optimal generalization of   the Bell inequality  in the case of a bipartite system made of two qutrits is the Collins, Gisin, Linden, Massar and Popescu (CGLMP) inequality~\cite{CGLMP,CGLMP2}.
In order to explicitly write this condition, consider again the components $S_{A}$ and $S_{B}$ of the bipartite qutrit system. For the qutrit $S_{A}$, select two spin measurement settings, $\hat{S}_{A_{1}}$ and $\hat{S}_{A_{2}}$, which correspond to the projective measurement of two spin-1 observables having
each three possible outcomes $\{0,1,2\}$---that, in our case, take values in $ \{+1,\, 0,\,-1\}$. Similarly, the measurement settings and corresponding observables for the other qutrit $S_{B}$ are $\hat{S}_{B_{1}}$ and $\hat{S}_{B_{2}}$. Then, denote by $P(A_{i}=B_{j}+k)$
the probability that the outcome $S_{A_{i}}$ for the measurement of $\hat{S}_{A_{1}}$ and $S_{B_j}$ for the measurement of $\hat{S}_{B_{j}}$ , with $i$, $j$ either 1 or 2, differ by $k$ modulo 3. One can then construct the combination:
\bea
{\cal I}_{3}& =& P(A_{1}=B_{1}) +  P(B_{1}=A_{2}+1) +  P(A_{2}=B_{2})\nn \\
&+ &   P(B_{2}=A_{1})  -P(A_{1}=B_{1}-1) - P(A_{1}=B_{2}) \nn \\
& -& P(A_{2}=B_{2}-1)  -P(B_{2}=A_{1}-1)\ . 
\label{CGLMP}
\eea
For deterministic local models, this quantity satisfies the  generalized Bell inequality
\be
{\cal I}_{3}\leq 2\ ,
\label{CGLMP_inequality}
\ee
which instead can be violated by computing the above joint probabilities using the rules of quantum mechanics. 
\com{Given a state $\rho$ of the two-qutrit system, the above probabilities are computed in quantum mechanics
as expectation values of suitable projector operators; for instance, the probability of the outcome
$A_1=B_1=1$, when measuring $\hat{A}_1$ and $\hat{B}_1$, is given by 
$P(A_1=B_1=1)={\rm Tr}[\rho\, ({\cal P}_{A_1=1}\otimes{\cal P}_{B_1=1})]$, where {\it e.g.} ${\cal P}_{A_1=1}$
projects onto the subspace of the $A$-Hilbert space where $\hat{A}_1$ assumes the value~1. Therefore,} In quantum mechanics, ${\cal I}_{3}$ in \eq{CGLMP} can be expressed as an expectation value
of a suitable Bell operator $\cal B$ as 
\be
{\cal I}_{3}={\rm Tr}\big[\rho\, {\cal B}\big].
\ee
The explicit form of $\cal B$ depends on the choice of the four measured operators $\hat{A}_i$ and $\hat{B}_i$. Hence, given the two-qutrit state $\rho$, it is possible to enhance the violation of  the Bell inequality (\ref{CGLMP_inequality}) through a specific choice of these operators. The numerical value of the observable ${\cal I}_{3}$ is bound to be less than or equal to 4.
For the case of the maximally entangled state, the problem of finding an optimal choice of measurements has been solved~\cite{CGLMP}. By working in the single spin-1 basis formed by the eigenstates of the  spin operator in the direction $\hat z$ with eigenvalues $\{+1,0,-1\}$, the Bell operator takes a particular simple form:
\be
\small
{\cal B} =  \begin{pmatrix} 
  0 & 0 & 0 & 0 & 0 & 0 & 0 & 0 & 0  \\
  0 & 0 & 0 & -\frac{2}{\sqrt{3}} & 0 & 0& 0 & 0 & 0  \\
  0 & 0 &0 & 0 & -\frac{2}{\sqrt{3}} & 0 &2 & 0 & 0  \\
  0 &  -\frac{2}{\sqrt{3}} & 0 & 0 & 0 & 0 & 0 &0 & 0  \\
  0& 0 & -\frac{2}{\sqrt{3}} & 0 & 0 & 0 & -\frac{2}{\sqrt{3}} & 0 &0  \\
  0 & 0 & 0 & 0 & 0 & 0 & 0 &  -\frac{2}{\sqrt{3}} & 0  \\
  0 & 0 & 2 & 0 & -\frac{2}{\sqrt{3}} & 0 &0& 0 & 0  \\
  0 & 0 & 0 & 0 & 0 &  -\frac{2}{\sqrt{3}} & 0 & 0 & 0  \\
  0 & 0 & 0 & 0 & 0& 0 & 0 & 0 & 0  \\
\end{pmatrix} \ ,
\label{B}
\ee
 after rotating it into the helicity basis  from the so-called computational basis employed in~\cite{Acin:2002zz}.

Within the choice of measurements leading to the Bell operator, there is still the freedom of modifying the measured observables through local unitary transformations, which effectively corresponds to local changes of basis in the measurement of the polarizations. Correspondingly, the Bell operator undergoes the change:
\be
{\cal B} \to (U\otimes V)^{\dag} \cdot {\cal B}\cdot (U\otimes V)\ , \label{uni_rot}
\ee
where $U$ and $V$ are independent $3\times 3$ unitary matrices.
In the following we make use of this freedom to maximize the value of ${\cal I}_3$.

\vskip0.5cm
{\versal Results.---}  Our results can now be given in a very concise form. 
The polarization amplitudes in \eq{data0} determine the  polarization density matrix in \eq{rhoBVV} for the decay $B^0\to J/\psi \,\K$. The density matrix makes it possible to estimate the observables in which we are interested.

We determine the rotation matrices $U$ and $V$ in the optimization procedure of \eq{uni_rot} by means of the central values in \eq{data0}.
\com{ \be
 \small
U = \begin{pmatrix} 
  \dfrac{5}{14}-\dfrac{8 i}{13} & \dfrac{1}{20}-\dfrac{4 i}{7} & \dfrac{1}{4}-\dfrac{ i}{3}  \\\\
 0& \dfrac{1}{18}-\dfrac{7 i}{12} & -\dfrac{7}{15}+\dfrac{11 i}{17}  \\\\
 - \dfrac{4}{11}+\dfrac{3 i}{5} & \dfrac{1}{17}-\dfrac{4 i}{7} & \dfrac{1}{4}-\dfrac{i}{13}\\
\end{pmatrix} \, , ~~~
V = \begin{pmatrix} 
  -\dfrac{5}{12}-\dfrac{ i}{99} &-\dfrac{2}{9}+\dfrac{8 i}{15} & \dfrac{1}{51} + \dfrac{5 i}{7}  \\\\
  -\dfrac{4}{5}-\dfrac{i}{38} & \dfrac{3}{13}-\dfrac{6 i}{11}  &0\\\\
  -\dfrac{5}{12}-\dfrac{ i}{59} &-\dfrac{3}{13}+\dfrac{9 i}{17} &-\dfrac{5 i}{7}\\
\end{pmatrix} \, , ~~~
\label{UV}
\ee
the entry of which are written as rational numbers  (approximated at  1\%).}

We propagate the uncertainties in the polarization amplitudes in \eq{data0},  taking into account  also their correlations.
We find that the entropy of entanglement  among the polarizations of the final mesons for the decay $B^0\to J/\psi \, \K$ is  given by
\be
\mathscr{E} = 0.756 \pm  0.009\, . \label{sig_ent}
\ee
The result in \eq{sig_ent}  represents a detection of the presence  of quantum entanglement with a significance well above 5$\sigma$ (nominally 84$\sigma$).

Propagating the uncertainties through the expectation value of the Bell operator, while keeping the two matrices $U$ and $V$ fixed, we determine that ${\cal I}_3$  for the decay $B^0\to J/\psi \, \K$  has expectation  value
\be
{\cal I}_3= 2.548 \pm 0.015\, ,
\ee
and therefore the CGLMP inequality ${\cal I}_3<2$ is violated with a significance well above 5$\sigma$ (nominally 36$\sigma$).

Other decays of $B$ mesons provide polarization amplitudes  that can be used in  similar fashion to test   the Bell inequality. We list in Table~\ref{tab:others} the  values for the entanglement $\mathscr{E}$ and  the Bell operator ${\cal I}_{3}$ for some of the   decays  we have considered. Specifically,  ${\cal I}_3<2$ is violated with a significance of  more than 5$\sigma$ in  the decays $B_s\to \phi \, \phi$ and $B_s\to J/\psi  \,\phi$.

\begin{widetext}
\begin{table*}
\begin{ruledtabular}
\begin{tabular}{l  c c}\\
  & $\boldsymbol{\mathscr{E}}$ &$\boldsymbol{{\cal I}_{3}}$ \\[+0.2cm]
\hline\\
$ \bullet\; \;\boldsymbol{B^{0}\to J/\psi  \,\K}$ \cite{LHCb:2013vga}\hspace{1cm} & $\qquad 0.756 \pm 0.009 \qquad$   & $2.548 \pm 0.015$ \\[+0.2cm]
$\bullet\;\;\boldsymbol{B^0\to \phi \,\K}$ \cite{Belle:2005lvd} & $0.707\pm 0.133^{\ast}$    &  $2.417\pm0.368^{\ast}$\\[+0.2cm]
$\bullet\;\;\boldsymbol{B^{0}\to \rho \, \K }$ \cite{LHCb:2018hsm} &  $0.450\pm 0.077^{\ast}$  &  $2.208 \pm 0.151^{\ast}$\\[+0.2cm]
$\bullet\;\;\boldsymbol{B_s\to \phi \, \phi}$ \cite{LHCb:2023exl} & $0.734\pm 0.037$    &  $2.525\pm 0.064 $\\[+0.2cm]
$\bullet\;\;\boldsymbol{B_s\to J/\psi  \,\phi}$ \cite{ATLAS:2020lbz} & $0.731\pm0.032$    &  $2.462\pm 0.080 $\\
\\
%\hline\\%
\end{tabular}
\caption{\footnotesize \label{tab:others} \textrm{Entanglement and Bell operator ${\cal I}_3$  for some $B$-meson decays. An asterisk indicates that  the correlations in the uncertainties of the helicity amplitudes are not  given in the corresponding reference  and  therefore only an upper bound on the  propagated uncertainty can  be  computed.}}
\end{ruledtabular}
\end{table*}
\end{widetext}

\vskip0.5cm
{\versal  Outlook.---} We have shown that quantum entanglement is present and  the  Bell inequality is violated by the data on the polarization amplitudes in the decay $B^0\to J/\psi \, \K$, and other similar decays. The presence of entanglement and the Bell inequality violation have  very large significances and establish  these property of quantum mechanics  at   high energies in a collider setting and in a system in which all  Standard Model interactions are involved. It is the first time that the violation of the inequality is shown to take place in a system of two qutrits and between two different particles.

We are aware that potential loopholes are present in any test of the Bell inequality. These loopholes have been closed in low-energy tests with photons~\cite{Giustina:2015yza,Hensen:2015ccp}
and in atomic physics~\cite{Rosenfeld}.

To close the {\it locality loophole}---which exploits events not separated
by a space-like interval, as it is the case of the $J/\psi\, K^*$ decays---one must consider decays in which the produced particles are identical, as in the $B_s\to \phi\phi$ decay, and therefore their life-times are also the same. The actual decays take place with an exponential spread, with,  in the $\phi\phi$ case, more than 90\% of the events being separated by a space-like interval. 

The presence and relevance of other possible loopholes  is still an open question in the high-energy setting and beyond the scope of the present Letter.
 
\vskip0.5cm
\textit{Acknowledgements.---} We thank M. Dorigo for discussions on various aspects of $B$-meson physics. L.M. is supported by the Estonian Research Council grant PRG356.
 
%%%%%%%%%%%%%%%%%%%%%%%%%%%%%%%%%%%%%

%%%%%%%%%%%%%%%%%%%%%%%%%%%%%%%%%%%%%%%%%%%%%%

\end{document}